\documentstyle[epsf]{article}
\def\Journal#1#2#3#4{{#1} {\bf #2}, #3 (#4)}

\def\NPB{{\em Nucl. Phys.} B}
\def\NPA{{\em Nucl. Phys.} A}
\def\PLB{{\em Phys. Lett.} B}

\def\PR{{\em Phys. Rep.} }
\def\PRL{{\em Phys. Rev. Lett.}}
\def\PRD{{\em Phys. Rev.} D}
\def\ZPC{{\em Z. Phys.} C}

\newcommand{\ds}{\displaystyle}

\newcommand{\unit}[2]{$#1\,${\rm #2}}
\newcommand{\fmc}[1]{$#1\,{\rm fm}/c$}
\newcommand{\betrag}[1]{\mbox{$\left|#1\right|$}}

\newcommand{\vek}[1]{{\bf #1}}

\begin{document}
\begin{center}
{\Large \bf Nonequilibrium dynamics of a hadronizing 
quark-gluon plasma\footnote{This work was supported by BMBF, DAAD, DFG, GSI, Graduiertenkolleg 
Theoretische und Experimentelle Schwerionenphysik, and
the A.~v.~Humboldt Foundation.}
}
\end{center}
\vspace{0.3cm}

\begin{center}

{M.~Hofmann$^a$, 
J.~M.~Eisenberg$^b$, 
S.~Scherer$^a$, 
M.~Bleicher$^{a,}$\footnote{Fellow of the Josef Buchmann foundation}, 
L.~Neise$^a$, 
H.~St\"ocker$^a$,
W.~Greiner$^a$} 
\vspace{0.3cm}

{\it $^a$Institut f\"ur Theoretische Physik, J.~W.~Goethe-Universit\"at,\\ 
D-60054 Frankfurt am Main, Germany},\\
{\it $^b$School of Physics and Astronomy, Tel Aviv University,\\
69978 Tel Aviv, Israel } 
\end{center}

\vspace{0.5cm}

{\bf Abstract}
\vspace{0.5cm}

We investigate the hadronic cooling of a quark droplet within 
a microscopic model. The color flux tube approach is used
to describe the hadronization of the quark phase. The model 
reproduces experimental particle ratios equally well compared to a
static thermal hadronic source. Furthermore, the dynamics of the
decomposition of a quark-gluon plasma is investigated and time dependent particle
ratios are found.

\vspace{1.0cm}

%
%

\section{Introduction}

When the possible existence of a quark-gluon plasma was first
postulated in the mid seventies \cite{Collins:Polyakov} 
several groups started to investigate unique signatures which 
would allow the identification of such a gas of asymptotically free 
quarks and gluons in heavy ion collisions. 
Tightly connected to this task was the study of the mechanism 
for the decomposition of a once formed partonic state of matter, 
back to the hadronic world measured in detectors. One was immediately 
lead to the question of the order and signatures of 
a QGP-hadron phase transition. 
Under the widely used assumption of a first order transition
a sharp surface of the quark phase could be defined
at which the hadronization should take place. 
Applying Gibbs conditions at the phase boundary 
then gives a macroscopic view on the properties 
of the initial quark and the final hadron phase. 
However, the dynamical aspects of the plasma disassembly 
(e.~g.\ the dissociation rate) could not be tackled 
in this simple picture. 

Therefore, several different models for the dynamics of the 
transition were proposed which all focused on the  
radiation of hadrons from the plasma surface. 
One of these contributions was driven by our late friend
Judah Eisenberg \cite{MUELL84} who proposed a quark-gluon bag which
was surrounded by a cloud of pions. Both phases interact at the phase
boundary via the coupling of the pseudo-scalar pion field to the
quark vector-field in the interior of the bag. 
However, this and other similar approaches as e.g. \cite{DANOS83} 
still lacked a dynamical description of the conversion process 
and therefore seemed to be restricted to stationary problems in equilibrium. 
This problem was partially attacked by the application 
of the color flux tube model to the radiation scenario 
as proposed in \cite{BANER83}. 
Here, the idea of confinement is adopted 
to describe hadronization dynamics. Now a dynamical
non-equilibrium calculation of the pion production rates was
possible. However, all those approaches predicted that the
contribution of surface radiation to the total hadron yield 
in heavy ion collisions should be small compared to the strong
hydrodynamic expansion of a hot fireball. 
Even the extension of the flux tube model to four quark flavors and
the complete hadron zoo could not circumvent this result. 

Nevertheless, the hadron radiation process is yet not
completely negligible which is underlined by the rough accordance of
all those different models within one order of magnitude
(fig.~\ref{figure1}). But, evidently, the dynamics of the phase
transition cannot be separated from the dynamics of the QGP phase
itself which includes the radiation from the surface as well as the
expansion. The latter, however, requires a new, dynamical definition
of the ``surface'' of a QGP as this surface will now be time
dependent. In the microscopic picture of moving quarks (as the flux
tube model) a surface cannot be defined consistently: it has to be
introduced ``by hand'' as a boundary condition. The complete
microscopic description will therefore not contain such a
surface. Inspired by Judah Eisenberg we extended the flux
tube approach by assuming the confining force acting on an escaping
quark as dynamically induced by all the other quarks which define
the plasma phase, i.~e. one quark moves in the effective color field
of all the others. 

This paper presents this model to describe the dynamics of the
hadronization transition. It explicitly takes into account a
non-equilibrium evaporation scenario.

\section{Hadron radiation from quark-gluon plasma}

Two fundamental scenarios have been suggested for the decay of
a quark-gluon plasma within the past fifteen
years. The first scenario assumes a hydrodynamical expansion of the plasma 
and a cooling  which  leads to hadronization \cite{BJORK83}.
In a second approach hadronization is constrained to the plasma
surface mediated by a particular interaction between the two phases.
Several hypotheses have been developed, all of which assume 
an equilibrated  plasma state that evaporates mesons according to specific
mechanisms. Reciprocally, the plasma should also be able to
absorb hadrons from outside with a certain probability. 

The pion emission rate of a thermalized quark gas
has been calculated \cite{DANOS83} by introducing 
a fixed threshold quark momentum for the hadronization of escaping
quarks to pions. Thus the pion yield is determined by the characteristics 
of the initial plasma state, and the energy flux of pions per unit time 
and surface element ${\rm d}^3E/{\rm d}^2S\,{\rm d}t$
obeys a Stefan-Boltzmann $T^4$ law.

A more sophisticated mechanism of meson evaporation from a plasma state
uses the cloudy-bag model postulating the existence of a pseudo scalar
pion field outside the plasma region coupled to the quarks at the
surface \cite{MUELL84}. In this model pions are produced  by the processes
$q\to q\pi$, $\bar q\to\bar {q}\pi$ and $q\bar{q}\to\pi$ at the
surface of the bag. The result of this model differs drastically 
from the predictions of \cite{DANOS83}, revealing an approximate 
power law for the emissivity of $T^6$.

In this paper we apply a hadronization scheme 
proposed by Banerjee et al. \cite{BANER83}. A quark penetrating the
bag surface is pulled back by a strong color interaction with
the volume of the  plasma. The confining force  increases with the
distance from the plasma so that a quark cannot escape. This
attractive potential is simulated by a color flux tube connecting
the emitted quark to the plasma by a uniform color field. This tube
can decay by creation of virtual $q\bar{q}$ pairs inside the color field. 
The color charges of the pair produce screening of
the color field between the initially emitted quark 
and the quark droplet and allow $q$ and $\overline{q}$ to build a
color singlet state (a meson). The remaining $q$ will be trapped in a
color flux tube again. 

Baryon production can be implemented e.~g. by
inclusion of  (anti-)diquark states ($uu$, $ud$, $\ldots$) 
which later may combine with a single quark to form a baryon.
However, the probability for $qq-\overline{qq}$ production 
is strongly suppressed  by the relatively large 
masses of the diquark states, so that the predicted baryon
multiplicities underestimate the experimental result.
Therefore, our calculation treats diquarks as constituents 
of the quark-gluon plasma. This mechanism will drastically increase 
the number of emitted baryons.

The existence of bound diquarks states in the QGP conflicts with 
the expectations of an asymptotically free quark-gluon plasma.
However, recent calculations of several groups hint at
the existence of correlations in the QGP even for $T\le 2T_C$ . 

We assume that a quark-gluon plasma bag has a radius of  
$R\sim 2-6\, {\rm fm}$ and is in complete
thermal and chemical equilibrium at  temperatures 
$T\sim 100-250\,{\rm MeV}$. It is filled with asymptotically 
free quarks and antiquarks.
In the following we neglect the gluonic degrees of freedom. 

Let us consider a thermal quark plasma bag with a sharp surface. 
This state will then radiate hadrons from the surface which 
change the energy and flavor content of the
plasma. We consider that near the surface the particles are homogeneously
distributed in space, while their momentum distribution 
is approximated by the Boltzmann distribution $\exp(-\sqrt{p^2+m_q^2}/T)$.
We assume to  have current quarks (and diquarks) inside the plasma. 
In this approximation the mean particle number for 
each species $q$ inside the bag is 
\begin{equation}\label{number}
N_q(T,V,\mu) = g_q\,e^{\mu/T}\,\frac{V}{2\pi^2}m_q^2\,T\,K_2\left(\frac{m_q}{T}\right)\;,
\end{equation}
where $q$ runs over all possible (anti-)quark and (anti-)diquark
states, and $g_q$ is the degeneracy factor which counts all possible
color and spin states. 

In contrast to the other models of hadron radiation cited in this article, 
we are, in principle, {\it not} restricted to a thermal initialization 
of the plasma phase.

\subsection{Hadronization via fission of color flux tubes}

The  string  mechanism proposed in
\cite{ANDER78} describes the fragmentation of a color flux tube 
created between two point-like quarks excited by a high impact
$p+p$ (or heavy ion) collision. It is assumed that longitudinal
momenta of quarks are much higher than the transverse components 
so that a  $1+1$ dimensional description is appropriate. 
A string stretched between a quark and a large
bag is different since the color field is not necessarily constrained to
a tube. However, assuming free motion of color charges 
on the surface of the bag, we assume a flux tube which is 
perpendicular to the bag surface and will follow a possible quark motion
perpendicular to the tube. 

A quark crossing the surface of the bag travels outward in
accordance with the attractive potential of the color flux tube. 
In $1+1$ dimensions this potential is
taken to be linear
\begin{equation}
	V(z) = \kappa\cdot z\;,
\end{equation} 
which assumes 
 a constant chromoelectric field strength density ${\rm
d}V/{\rm d}z = \kappa\approx 0.9\,{\rm GeV/fm}$. 
This value can be extracted from fits to
Regge trajectories  or from lattice calculations \cite{BORN94}.

The total probability of vacuum decay is then given by
\begin{equation}\label{probflav}
p = \sum\limits_{f = q,qq}\,p_f\, , \qquad
p_f = \frac{\kappa}{4\pi^3}\,\sum\limits_{n=1}^\infty\,
\frac{1}{n^2}\,e^{\displaystyle -\frac{\pi m_f^2\,n}{\kappa}}\;,
\end{equation}
where the sum runs over all quarks and diquarks, 
$p_f$ is the probability per unit four-volume to produce a 
virtual $q_f\bar{q_f}$ pair in a constant field \cite{CASHER79,GLEND83} ,
and $m_f$ is the mass of the produced quarks (or diquarks). 
The total probability of vacuum decay is then given by
This result has the same form as  Schwinger's formula  for 
$e^+e^-$ production in an constant external field \cite{SCHWING51}. 
Nevertheless, it should be noted that the above result is valid only
for a static, homogeneous color field, i.~e. for a
flux tube with infinite  transverse dimensions. Finite  flux tubes
impose  surface effects which may noticeably modify  the field
configuration and pair production probabilities. In \cite{SCHOEN90} it
was shown that a string radius of $0.5\,{\rm fm}$ reduces the pair
creation rate by about 40\%.

For the quark masses we assume the following values of $m_u = m_d = 5\, {\rm
MeV}, m_s = 150\, {\rm MeV}, m_c = 1.5\, {\rm GeV}$. For the mass of the
diquarks two different possibilities are taken into account: masses derived
from fits to experimental $e^+e^-$ data which allow to extract a
probability ratio $p(qq)/p(q) = 0.065$ for diquark production \cite{Bel82},
and, secondly, diquark masses calculated within the 
Nambu-Jona-Lasinio model \cite{VOGEL91}. One obtains different masses for 
different flavor and spin combinations. 
Combining $u$, $d$ and $s$ flavors and using simple symmetry arguments
for the diquark wave functions (see e.g.\cite{SZCZ89}) one obtains three antisymmetric
diquark states of spin 0 
\begin{equation}
(ud)_0 = \frac{1}{\sqrt{2}}(ud-du),\quad
(us)_0 = \frac{1}{\sqrt{2}}(us-su),\quad
(ds)_0 = \frac{1}{\sqrt{2}}(ds-sd)\;,
\end{equation}
and six symmetric combinations
\begin{eqnarray}
(uu)_1 &=& uu,\quad (ud)_1 = \frac{1}{\sqrt{2}}(ud+du),\quad
(dd)_1 = dd,\\
(ss)_1 &=& ss,\quad(us)_1 =
\frac{1}{\sqrt{2}}(us+su),\quad(ds)_1 =
\frac{1}{\sqrt{2}}(ds+sd)\;,
\end{eqnarray}
with spin 1. Since only the anti-triplet  representation of the
$SU(3)_c\times SU(3)_c$ may contribute to a  diquark state, we
obtain a color degeneracy of $3$. Considering all the spin and
isospin combinations, we arrive at the  total degeneracy
factors for each kind of diquarks given in table \ref{tab1}.

\begin{table}[tb]
\begin{small}
\tabcolsep0.3em
\renewcommand{\arraystretch}{1.5}
\begin{tabular}{l|ccccc}
&$(qq)_0$,$(\bar q\bar q)_0$&$(qq)_1$,$(\bar q\bar q)_1$&$(qs)_0$,$(\bar q\bar s)_0$&$(qs)_1$,$(\bar q\bar s)_1$&$(ss)_1$,$(\bar s\bar s)_1$\\
\hline
LUND mass (MeV)&420&490&590&640&790\\
\hline
NJL mass (MeV)&234&822&537&962&1087\\
\hline
degeneracy&3&27&6&18&9\\
\end{tabular}
\end{small}
\caption{\small Diquark degeneracies and masses. Masses are  taken from fits
to the LUND string model and from calculations within the NJL model.}\label{tab1}
\end{table}

Assuming a string cross section $A$ and a constant break-up
probability $p$ inside  the string volume we obtain a time-dependent
probability for string fission
\begin{equation}\label{prob2}
P(t) = 1-e^{-p\,A\,\int\limits_0^{t}\,{\rm d}t'\,z(t')}
\end{equation}
where $z(t)$ is the distance of the quark from the surface. 
The flavor of the created quarks and
therefore the species of the hadron to be produced is chosen according to
the probability given by (\ref{probflav}), defining all quantum numbers. 
If this choice proves to be energetically 
impossible the produced flavor will be reselected.
The remaining quark is propagated within the attractive color 
potential from  the position of the point of break-up until the
flux tube breaks up again or the quark is pulled back into the plasma.
String fission is only possible while the quark is moving outward ($p_q^z>0$). 

The flavor and the quantum numbers of the produced $q\overline{q}$
(or $qq-\overline{q}\overline{q}$)
pair determine 
the species of the produced hadron. Quantum mechanical binding
properties have been neglected as 
well as possible mass reduction of the hadrons due to the dense
surrounding medium. Thus the free vacuum mass for the hadrons is assumed.
The probability for the $q\overline{q}$
pair to have a 
transverse momentum $p_\perp$ is
\begin{equation}
	P(p_\perp) = e^{\ds -\frac{\pi
E_\perp^2}{\kappa}}\qquad\mbox{with}\quad
E_\perp = \sqrt{M^2+(P_M^\perp)^2}\;.
\end{equation}
From momentum conservation during the string fission we obtain 
directly the transverse momentum of the meson. The
longitudinal hadron momentum is determined by using the fragmentation
function $f$ according to \cite{FIELD77,ANDER78,ANDER83}
\begin{equation}\label{fragfun}
z\,f(z) \sim (1-z)^\alpha\,e^{-\beta\,E_\perp^2/z}\;,
\end{equation}
where $z = p_{\rm
hadron}/p_{\rm initial\, quark}$ and $\alpha,\beta$ are free
parameters which are fixed in the LUND model to $\alpha = 1$ and
$\beta = 0.7\,{\rm GeV}^{-2}$.

\subsection{Results}

\begin{figure}[tb]
\centerline{\parbox[b]{7cm}{\epsfxsize=7cm \epsfbox{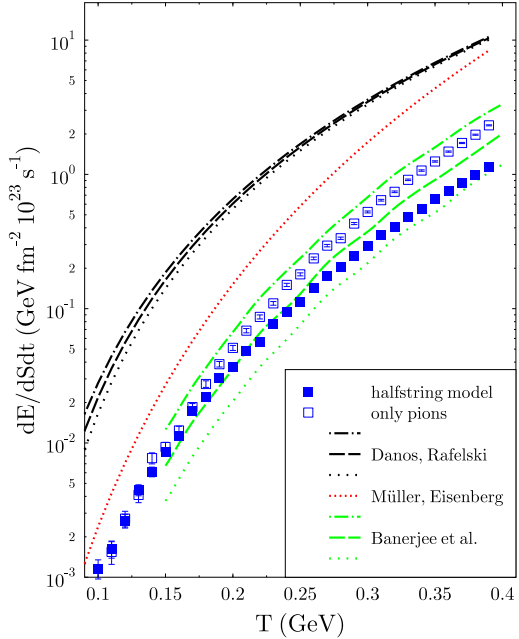}}\hfill
\parbox[b]{4.3cm}{\caption{\small Pion surface brightness of a baryon-free QGP as a function of
temperature. Three upper lines are the results from Danos and
Rafelski \protect\cite{DANOS83}, the dotted line are  calculated by
M\"uller and Eisenberg \protect\cite{MUELL84} while the three lower lines
represents the results from Banerjee et al. \protect\cite{BANER83} for various
model parameters. The filled squares are the calculations with
inclusion of all hadrons species while the open squares
give the results, when only pion production is taken into account.}\label{figure1}}}
\end{figure}

Figure~\ref{figure1} shows the surface brightness 
${\rm d}^3E_\pi/{\rm d}S\,{\rm d}t$ of the pion radiation 
as function of temperature $T$. We compare predictions of 
three different models by Danos and Rafelski \cite{DANOS83} (upper
lines), M\"uller and Eisenberg \cite{MUELL84} (dotted line) and Banerjee et
al. \cite{BANER83} (lower lines) with our results. We
predict the same orders of magnitude as in \cite{BANER83}.
The bending of the curve increases heavily
with increasing temperature which is due  to the opening of new hadronic
degrees of freedom  at higher energies. To see this, 
we recalculated the distribution neglecting all hadrons other 
than pions ($\Box$) which then gives a good agreement to the former results.


As describes so far, our model assumes a static plasma source. It has been
shown in ref.~\cite{SPIELES98} that hadron ratios change with time, 
i.~e. during the reaction some particle species are produced earlier than others. 
Furthermore, the hadron evaporation in each time step will influence 
the statistical  properties of both  the plasma phase and the hadronic phase. 
This dynamical evolution will pull the system from an initial $f_s=0$ 
into a domain of finite strangeness in both phases \cite{SPIELES98}. 


\begin{figure}[tb]
\centerline{\parbox[b]{6cm}{\epsfxsize=6cm \epsfbox{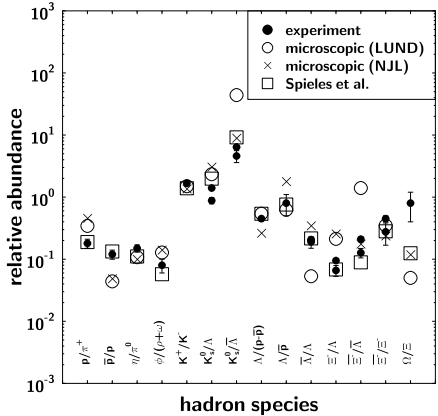}}\hfill
\parbox[b]{5.3cm}{\caption{\small Final particle ratios in the non-equilibrium scenario with initial
conditions $A_B^{\rm init} = 
100$, $S/A^{\rm init} = 45$, $f_s^{\rm init} = 0$, $B^{1/4} =
235\,{\rm MeV}$ corresponding to the calculations in
\protect\cite{SPIELES98} (squares). The circles
represent the results assuming LUND diquark masses
\protect\cite{ANDER83}, while the crosses 
denote calculations for diquark masses extracted from the NJL model
\protect\cite{VOGEL91}. 
Data from various experiments as compiled in 
\protect\cite{PBM} 
are also shown.}\label{figure2}}}
\end{figure}

It is interesting that the model  is  able to
reproduce experimental data fairly good even in a static approach, as can be
seen in fig.~\ref{figure2}. The initial conditions of the plasma phase
have been chosen to be equivalent to those used in \cite{SPIELES98}
$A_B^{\rm init} = 
100$, $S/A^{\rm init} = 45$, $f_s^{\rm init} = 0$, $B^{1/4} =
235\,{\rm MeV}$.
The NJL diquark masses seem to reproduce the
experimental data better than LUND masses. This is not surprising,
since the LUND masses have been extracted from fits of the string
model to high energy data ($\sqrt{s}\approx 4\,{\rm GeV}$) while 
average quark momenta in our scenario are certainly below 1 GeV.

One sees that our calculations fit the data almost as well
as the dynamical model and are in accordance to a static
thermal fit. This is ensured by the inclusion of diquarks 
in our plasma. First,  they are responsible
for creating baryon numbers but  secondly  they
reduce the distill effect of net strangeness. Indeed, at finite
$\mu_q$ an excess
of $q$ compared to $\overline{q}$ leads to enhancement of
$\overline{s}$ and therefore $K^+$ in the hadronic phase. The  
$qq$ and $qs$ diquarks also exceed the number of $\overline{qq}$ and
$\overline{qs}$ (suppressed) diquarks, but the latter may now
combine with a $s$ to form a $\Lambda$, $\Sigma$ or $\Xi$. This then
leads to an enhancement of $s$ in the hadronic phase. The net
strangeness enrichment in the hadronic sector is therefore 
balanced by the diquark contribution which is mostly influenced by
the initial baryon chemical potential $\mu_q$. 

\begin{figure}[tbp]
\centerline{\mbox{\epsfxsize=5cm \epsfbox{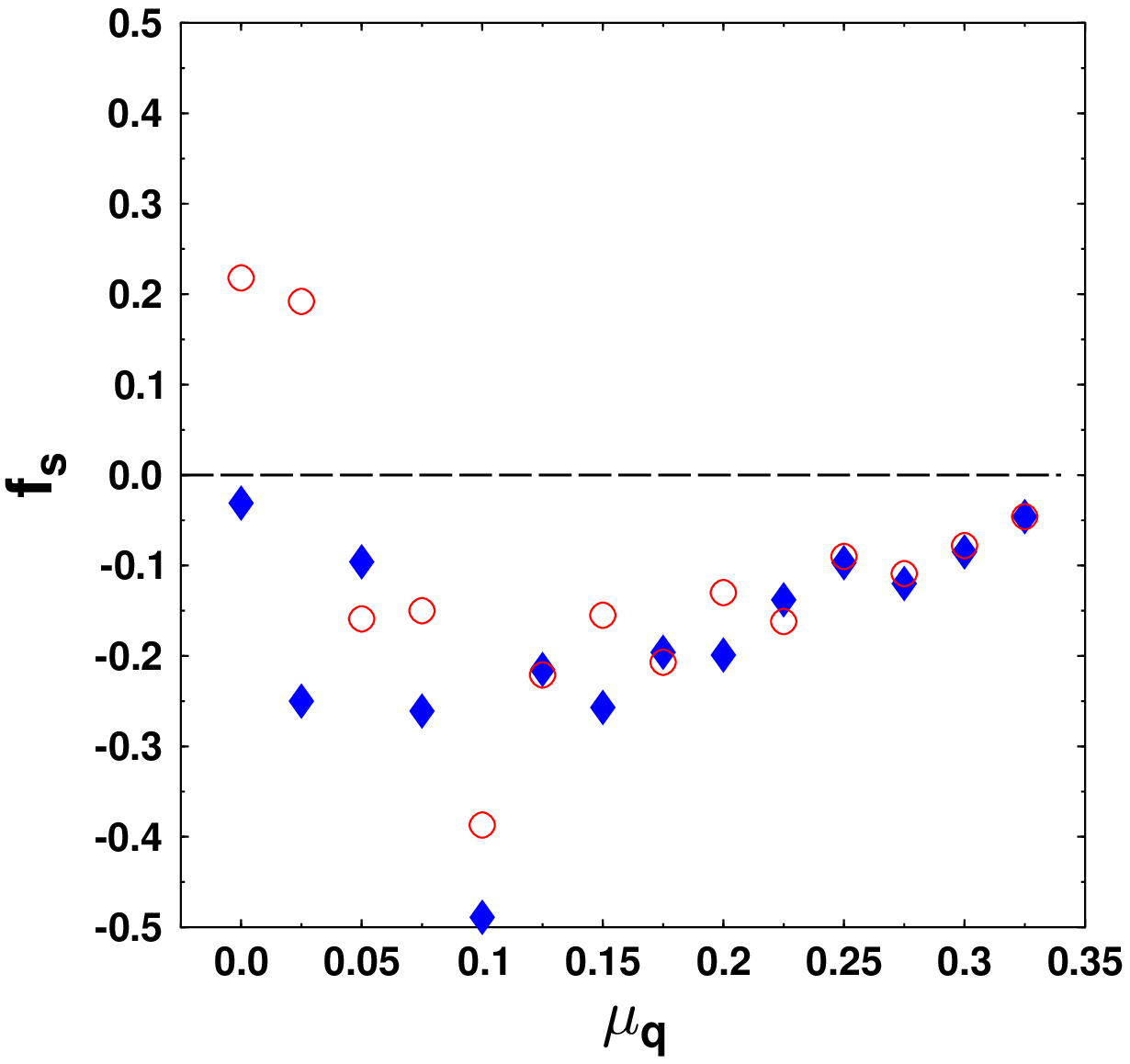}} \hfill \mbox{\epsfxsize=5cm \epsfbox{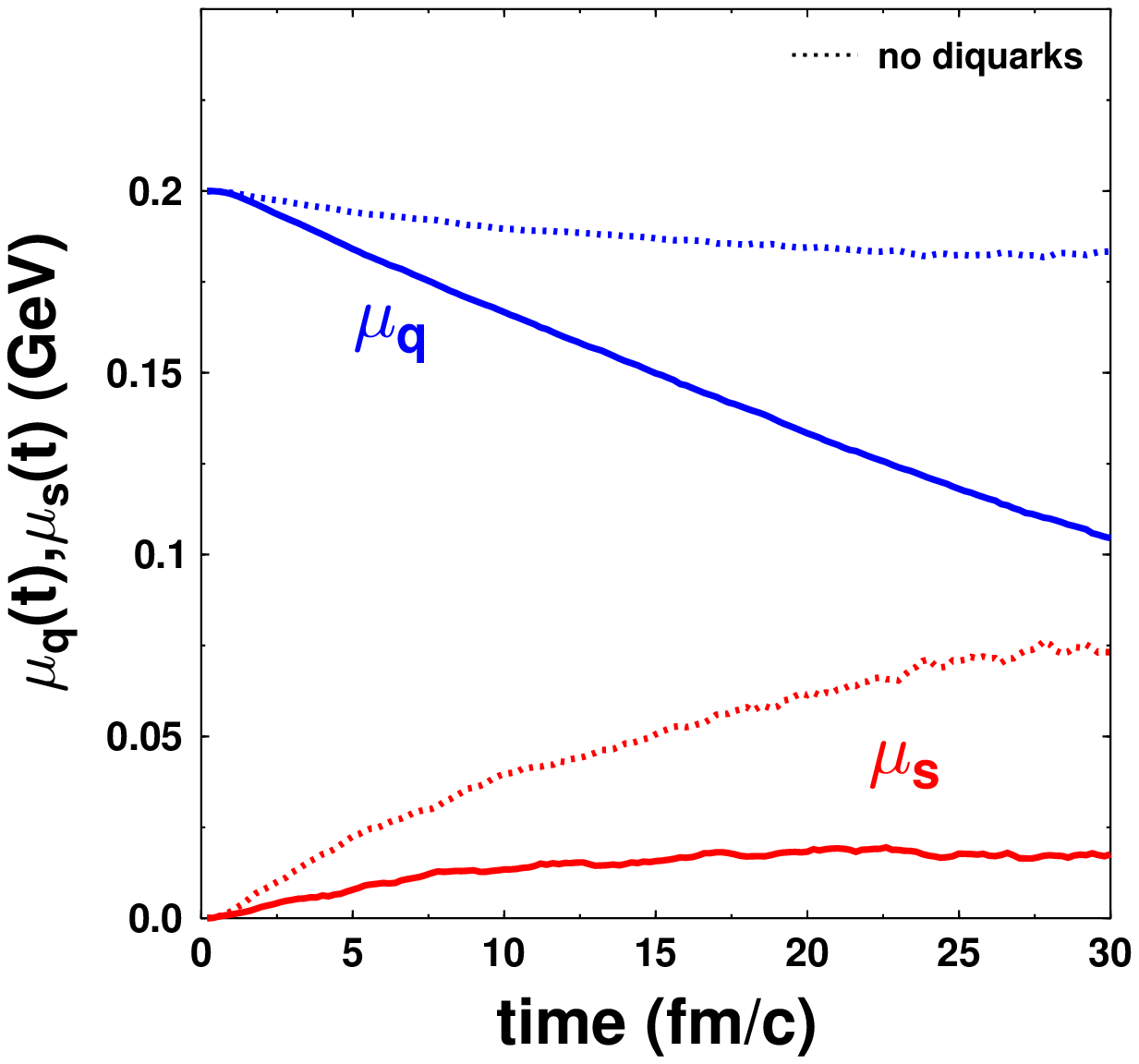}} }
\vspace{0.1cm}
\caption{\small (Left) Strangeness fraction $f_s = (N_s-N_{\overline{s}})/A$ as a
function of initial baryon chemical potential $\mu_q$ from microscopic
calculation for $\mu_s^{\rm init} = 0$ ($\Diamond$) and $\mu_s^{\rm
init} = 25\,{\rm MeV}$ ($\bigcirc$). (Right) Time evolution of chemical 
potentials $\mu_q$ and $\mu_s$ assuming a totally equilibrated plasma 
phase at every stage of the reaction with an initial temperature $T=190\,{\rm MeV}$. 
The dashed lines give the respective $\mu$'s for a  
plasma without diquarks.
}
\label{figure3}
\end{figure}

This is illustrated in
fig.~\ref{figure3} where the strangeness fraction $f_s =
(N_s-N_{\overline{s}})/A$ is plotted as 
a function of the chemical potential $\mu_q$ for $\mu_s=0$ (diamonds)
and $\mu_s=25\,{\rm MeV}$ (circles). For small $\mu_q$ diquark
production in the plasma is suppressed due to the high diquark masses and
therefore $\overline{s}$ is enriched on the hadronic phase giving rise
to negative $f_s$. For higher $\mu_q$ the strangeness fraction
goes to zero 
which indicates that both effects almost cancel. At $\mu_q\to 0$ there
should be $f_s=0$ due to restoration of  baryon number symmetry.
From the microscopic calculation it is not clear whether there is a
smooth transition in $f_s$ from finite $\mu_q$ to $\mu_q=0$ or a
discontinuity as net strangeness and baryon numbers 
converge to zero. The large deviation of two points at small $\mu_q$ indicate a
significant uncertainty in this observable.

For small but finite initial $\mu_s$ we observe a positive $f_s$ for
small $\mu_q$, as it can be expected. For higher $\mu_q$
the described $\overline{s}$ transport mechanism becomes more important and
finally dominates the initial strangeness excess. 

From fig.~\ref{figure3} (left) it is obvious that one feature of
a dynamical treatment of hadronization cannot be described in a static
scenario: the accumulation of anti-strangeness in the hadronic phase
causes an excess of strangeness inside the plasma which will then
change the strange chemical potential to finite $\mu_s>0$. The energy
stuffed into strangeness will then cause a strong reduction of
$\mu_q$. Figure \ref{figure3} (left) shows that for $\mu_q>50$ MeV this will
cause an even higher strangeness enrichment. For small $\mu_q$, on the
other hand, the 
growing $\mu_s$ will finally pull the system in a domain of positive
$f_s$, which means $s$-dominance in the hadronic phase which can no
longer be compensated by the small baryon chemical potential. 

\begin{figure}[tb]
\centerline{\epsfxsize=9cm \epsfbox{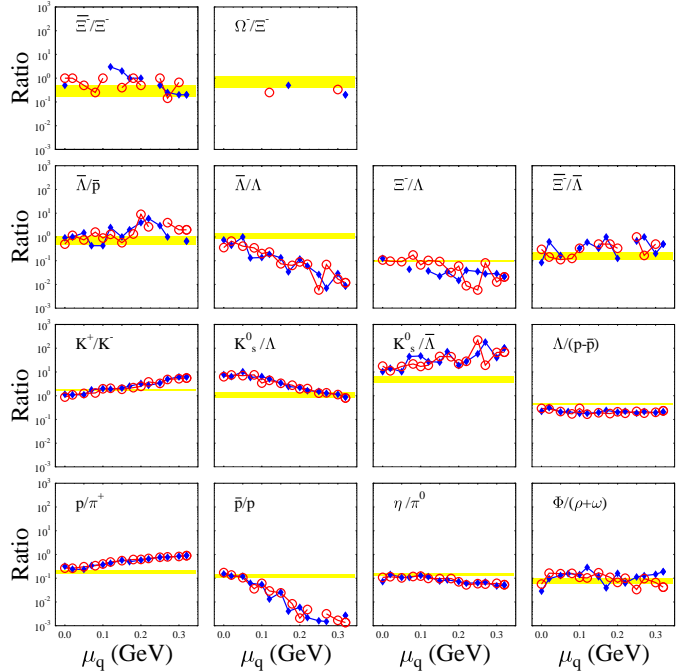}}
\vspace{0.5cm}
\caption{\small Hadron ratios as a function of $\mu_q$ for $\mu_s = 0$
($\Diamond$) and $\mu_s = 25\,{\rm MeV}$ ($\bigcirc$). The grey bars
indicate the experimental data including error bars.}
\label{figure5}
\end{figure}

Fig.~\ref{figure3} (right) elucidates this scenario. The time evolution of
$\mu_s$ and $\mu_q$ is plotted for initial conditions $T=190\,{\rm
MeV}$, $\mu_q^{\rm init} = 200\,{\rm MeV}$ and $\mu_s^{\rm init} =
0$ for both a plasma with (solid lines) and without (dotted lines)
initial diquarks. The time dependence is obtained by adjusting the
thermodynamical 
conditions inside the plasma to the average baryon number,
strangeness and energy transport 
through the plasma surface calculated by the microscopic model for
given initial conditions. The assumption of  a complete statistical equilibrium
of the plasma after each time step allows
to extract new parameters $\mu_q$, $\mu_s$ (as we are treating only
particle rates the volume $V$ of the plasma is of no importance for
particle ratios, unlike the absolute number of produced hadrons which is
proportional to the plasma surface). The temperature is assumed to be
constant. It is worth to note that in this scenario we do not need to
require equilibration 
for the hadronic phase or validity of Gibbs conditions at the 
transition. Thus, 
our model is a non-equilibrium description of the hadronization procedure.

In absence of diquarks in the plasma the strange chemical
potential monotonically rises with  time. This is in agreement to the
results of ref.~\cite{SPIELES98}. If diquarks are taken into account the
picture changes drastically. After a short period the $\mu_s$
contribution saturates at about 25 MeV. On the contrary the $\mu_q$
distribution  decreases linearly in time which gives possibility to
use  $\mu_q$ as  a time
evolution parameter. The strangeness distillery is strongly reduced by
including diquarks.

Finally, fig.\ \ref{figure5} shows the $\mu_q$ dependence  of different
particle ratios compared with experimental data for $\mu_s = 0$
and $\mu_s = 25\,{\rm MeV}$, where the latter represents the
saturation value for 
$\mu_s$ in the time dependence. Within the statistical fluctuations of
the ratios, the two curves cannot be distinguished. Therefore, the
distributions may be viewed as some kind of time evolution of the
particle ratios.

\section{The quark molecular dynamics (qMD)}

A novel extension of the quark dynamics can be realized by treating 
quarks as semi-classical particles which interact according to a two-body 
color potential (for details see \cite{Hof99}). This potential is phenomenologically 
motivated in order to mimic  the soft gluonic part of a quark-gluon plasma. 
The Hamiltonian reads
\[
{\mathcal{H}} = \sum_{i=1}^N\sqrt{p_i^2+m_i^2}+\frac{1}{2}\sum_{i,j}C_{ij}^c V(\mbox{\betrag{\vek{r}_i-\vek{r}_j}})
\]
where $N$ is the number of quarks. Here we include four quark flavors ($u,d,s,c$) 
with current masses \unit{m_u=m_d=10}{MeV}, \unit{m_s=150}{MeV} and \unit{m_c=1.5}{GeV}.

$C_{ij}^c$ indicates the color factor which regulates the sign and
relative strength of the interaction between any two quarks and 
antiquarks depending on the color combination of each pair. 
They can easily be obtained from the quark-gluon vertex factors 
in color space and read
\[
C_{\alpha\beta}^c = \sum_{a=3,8}\lambda^a_{\alpha\alpha}\lambda^a_{\beta\beta}
\]
in the abelian approximation. This yields e.~g.\
\[
C_{RR}^c = -1 \,, \qquad C_{RG}^c = +\frac{1}{2} \,, \qquad 
C_{R\overline R}^c = +1 \,, \qquad C_{RG}^c = - \frac{1}{2}\,.
\]

It is worth to note that the relative strength of the above color
factors is rigorously enforced by  the requirement
of color neutrality of widely separated $q\overline{q}$ and $qqq$
states. 

To provide confining properties $V(r)$ is taken to be linear at
large distances. At short distances the strong coupling constant
$\alpha_s$ becomes small compared to $1$ which causes the one-gluon
exchange terms to dominate the interaction and therefore induces a
coulomb-like potential. In total, one obtains the well known Cornell-potential
\cite{EICH75}
\[
V(r) = -\frac{3}{4}\frac{\alpha_s}{r}+\kappa\,r\;.
\]
In the static case of infinite quark masses this interquark potential
has been confirmed by lattice calculations over a wide
range of quark distances \cite{BORN94}.

The second request to the model is to define a criterion how to
map those bound quark states to hadrons. Such a mechanism
is essential as the Hamiltonian is not tuned to describe bound and
truly confined hadron states. This has been done by the requirement 
that the total color interaction from a pair (or a three particle state) 
of quarks with the remaining system vanishes. 
Then, these $q\overline{q}$- and $qqq$-states do no longer
contribute to the color interaction of the quark gas. 
If a bound quark state fulfills the hadronization criterion it will
be mapped to an appropriate hadronic state with identical quantum
numbers. Spin and isospin of the hadron is randomly chosen according
to the probabilities given by the Clebsch-Gordon coefficients while 
the mass of the produced hadron is determined by energy
and momentum conservation. 

\subsection{Dynamics of the model}

We can now study the dynamical evolution of the system. We use
the Metropolis algorithm to generate a thermal distribution of 
quarks where longitudinally, a Bjorken-like velocity profile 
$v_z=z/t$ has been imposed. 

Within about \fmc{12} most of the initial quarks are completely
hadronized. It becomes obvious that this hadronization process is {\em
not} compatible to a radiating hadron source, but is ruptured into
several small quark blobs before finally dissolving. 

\begin{figure}[tb]
\centerline{\parbox[b]{7cm}{\epsfxsize=7cm \epsfbox{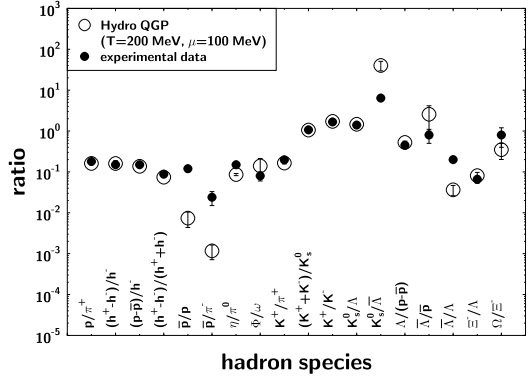}}\hfill
\parbox[b]{4.3cm}{\caption{\small Final state hadron ratios from qMD calculations (open
circles) compared to S+Au data at \unit{200}{$A$GeV} (full circles, taken from
\cite{PBM}).}\label{ratios}}}
\end{figure}

\begin{figure}[htb]
\centerline{\epsfxsize=9cm\epsfbox{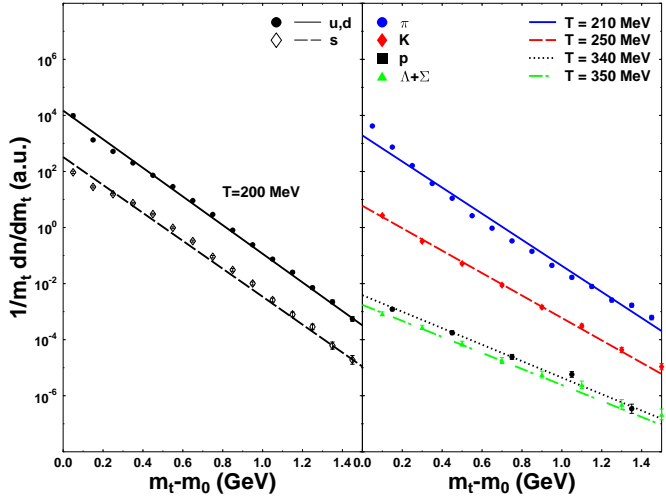}}
\vspace{0.1cm}
\caption{\small (Left) Transverse mass spectra of the initial
thermalized $u,d$ (circles) and $s$-quarks (open diamonds). Both
distributions show \unit{T=200}{MeV} as initialized.
(Right) Transverse mass spectra of  final $\pi$, $K$, $p$ and $\Lambda$ in
arbitrary units. One obtains rising inverse slope parameters with
increasing particles masses.}
\label{spectra}
\end{figure}

As the geometry of the quark phase resembles those of an assumed fireball in a
S+Au collision, we  compare in fig.~\ref{ratios} the calculated hadron
yields at mid-rapidity to those measured in CERN-SPS experiments 
(see compilation in \cite{PBM}).
We find a very good agreement in all $MM$, $MB$ and $BB$ ratios, while
the antibaryons seem to be clearly under-predicted. 
Particle ratios, however, proved not to be a very sensitive observable
to test the quality of theoretical models. Fits of a pure hadron gas
\cite{PBM} proved to describe data with a comparable precision as 
other thermal or hydrodynamical approaches including a QGP phase
transition or several microscopic simulations as UrQMD \cite{Bass}.
However, the analysis of event-by-event fluctuations \cite{Bleicher}
and of the dynamical properties of the system may yield new insight.
Thus, we shall further scrutinize the dynamical properties of the qMD
approach. As the initial conditions were taken from the idealized
Bjorken scenario we shall focus to the transverse dimension at
mid-rapidity. Fig.~\ref{spectra} (right) depicts the transverse mass 
spectra of various hadron species as produced by qMD with the above initial
conditions. The  distribution of the initialized quarks (left part)
shows \unit{T=200}{MeV} for all $u,d$ and $s$-quarks and underlines
the initialization procedure. In the final curves, however, we
observe a significant dependence of the slope of the distribution from
the mass of the regarded hadron species. We stress again that no
transversal flow had been imposed. This result is in strong contrast to
the fireball model proposed in \cite{Lee90} which explicitly required
a transverse velocity profile in order to obtain such transversal
behavior. 

\section{Conclusion}

Experimental particle ratios at SPS energies had been shown to be
compatible with the scenario of  of a static, thermally and chemically
equilibrated hadron gas. The present paper demonstrates that
the experimental data  are also compatible with a scenario with a  quark
gluon plasma consecutively evaporating hadrons, provided that 
the plasma contains thermally distributed bound diquark states.  
Without diquarks the model fails to describe the observed baryon 
abundances. Neither a bag constant nor equilibration
of the hadronic phase is assumed in this model.

Under the assumption that the plasma phase itself remains equilibrated 
during the reaction, a quasi time evolution of the plasma blob 
can be determined.
The model predicts strangeness distillation in the quark-hadron transition.
The inclusion of diquarks reduces the strangeness enhancement. 

The extension of the model to non-equilibrium initial conditions results
in a proper description of the transverse and longitudinal expansion
with a good agreement of the hadro-chemical properties.


%
%


\begin{thebibliography}{99}

\bibitem{Collins:Polyakov}
J.C.~Collins and M.J.~Perry,
\Journal{\PRL}{34}{1553}{1975} ;
A.M.~Polyakov,
\Journal{\PLB}{72}{224}{1977}

\bibitem{MUELL84} 
B.~M\"uller, J.~M.~Eisenberg,
\Journal{\PRL}{52}{1590}{1984}

\bibitem{DANOS83} 
M.~Danos, J.~Rafelski,
\Journal{\PRD}{27}{671}{1977}

\bibitem{BANER83} 
B.~Banerjee, N.~K.~Glendenning, T.~Matsui,
\Journal{\PLB}{127}{453}{1983}

\bibitem{BJORK83} 
J.~D. Bjorken,
\Journal{\PRD}{27}{140}{1983}

\bibitem{ANDER78} 
B. Andersson, G. Gustafson, C. Peterson,
\Journal{\NPB}{135}{273}{1978}

\bibitem{BORN94}
K.~D.~Born, E.~Laermann, R.~Sommer, T.~F.~Walsh, P.~M.~Zerwas,
\Journal{\PLB}{329}{325}{1994}

\bibitem{CASHER79} 
A. Casher, H. Neuberger, S. Nussinov,
\Journal{\PRD}{20}{179}{1979}

\bibitem{GLEND83} 
N.~K.~Glendenning, T.~Matsui,
\Journal{\PRD}{28}{2890}{1983}

\bibitem{SCHWING51} 
J. Schwinger,
\Journal{\PR}{82}{664}{1951}

\bibitem{SCHOEN90} 
T. Sch\"onfeld, A. Sch\"afer, B. M\"uller, K. Sailer, J. Reinhardt, W. Greiner,
\Journal{\PLB}{247}{5}{1990}

\bibitem{Bel82}
K.W.~Bell,  B.~Foster,  J.C.~Hart,  J.~Proudfoot,  D.H.~Saxon,  P.L.~Woodworth, 
Rutherford preprint {\bf  RL-82-011} (1982)

\bibitem{VOGEL91} 
U. Vogl,  W. Weise
\Journal{Prog.~Part.~Nucl.~Phys.}{27}{195}{1991}

\bibitem{SZCZ89} 
M. Szczekowski,
\Journal{Int.~J.~Mod.~Phys.~A}{4}{3985}{1989}

\bibitem{FIELD77} 
R.~D. Field, R.~P. Feynman,
\Journal{\PRD}{15}{2590}{1977}

\bibitem{ANDER83} 
B. Andersson, G. Gustafson, G. Ingelman, T. Sj\"ostrand,
\Journal{\PR}{97}{31}{1983}

\bibitem{SPIELES98} 
C.~Spieles, H.~St\"ocker, C.~Greiner
\Journal{Eur.~Phys.~J. C}{2}{351}{1998}

\bibitem{PBM} 
P.~Braun-Munzinger, J.~Stachel, J.P.~Wessels, N.~Xu,
\Journal{\PLB}{344}{43}{1995}; 
P.~Braun-Munzinger, J.~Stachel, J.P.~Wessels, N.~Xu,
\Journal{\PLB}{365}{1}{1996}; 
P.~Braun-Munzinger, J.~Stachel,
\Journal{\NPA}{606}{320}{1996}. 

\bibitem{Bass}
S.A.~Bass, M.~Belkacem, M.~Brandstetter, M.~Bleicher, L.~Gerland, J.~Konopka, 
L.~Neise, C.~Spieles, S.~Soff, H.~Weber, H.~Stocker, W.~Greiner
\Journal{\PRL}{81}{4092}{1998} 

\bibitem{Bleicher}
M.~Bleicher, M.~Belkacem, C.~Ernst, H.~Weber, L.~Gerland, C.~Spieles,
S.~A.~Bass, H.~St\"ocker, W.~Greiner
\Journal{\PLB}{435}{9}{1998} 

\bibitem{Hof99}
M.~Hofmann, M.~Bleicher, S.~Scherer, L.~Neise, H.~St\"ocker, W.~Greiner, submitted to {\em Phys. Lett.} B

\bibitem{EICH75}
E.~Eichten {\em et~al.},
\Journal{\PRL}{34}{369}{1975}

\bibitem{Lee90}
K.~S. Lee {\em et~al}., 
\Journal{\ZPC}{48}{525}{1990}




\end{thebibliography}
\end{document}